\documentclass{article}
\usepackage[utf8]{inputenc}
\usepackage{amsmath,amssymb}
\usepackage{bm}
\usepackage{graphicx}
\usepackage{hyperref}
\usepackage{xcolor}

\title{A Bayesian analysis of current duration data with reporting issues: an application to estimating the distribution of time-between-sex from time-since-last-sex data as collected in cross-sectional surveys in low- and middle-income countries}

\author{Chi Hyun Lee, Herbert Susmann, and Leontine Alkema \thanks{ The work was supported by the Bill \& Melinda Gates Foundation (INV-008441). Under the grant conditions of the Foundation, a Creative Commons 4.0 Generic License has already been assigned to the Author Accepted Manuscript version that might arise from this submission.} \footnote{Address for correspondence: Chi Hyun Lee, Department of Biostatistics \& Epidemiology, School of Public Health \& Health Sciences, University of Massachusetts Amherst, Amherst, MA 01003, USA. E-mail: \mbox{chihyunlee@umass.edu}}
\\
\bigskip 
\\Department of Biostatistics \& Epidemiology,\\School of Public Health \& Health Sciences,\\University of Massachusetts Amherst, Amherst, MA, USA. }

\date{}

\begin{document}

%%% just to make outline look ok:
%\parindent=0pt
%\setlength{\parskip}{1\baselineskip}
%%%

\maketitle
%\tableofcontents

\begin{abstract}
\noindent Aggregate measures of family planning are used to monitor demand for and usage of contraceptive methods in populations globally, for example as part of the FP2030 initiative. Family planning measures for low- and middle-income countries are typically based on data collected through cross-sectional household surveys. Recently proposed measures account for sexual activity through assessment of the distribution of time-between-sex (TBS) in the population of interest. 

% When calculating family planning measures from such surveys, sexual activity is taken into account only through checking whether a woman's most recent sexual intercourse occurred within a certain time period (e.g., the last four weeks). This approach does not fully utilize the information on the timing of sex. To develop new measures of family planning that better capture women's frequency of sex, a more  comprehensive assessment of sexual activity is needed. 

In this paper, we propose a statistical approach to estimate the distribution of TBS using data typically available in low- and middle-income countries, while addressing two major challenges. The first challenge is that timing of sex information is typically limited to women's time-since-last-sex (TSLS) data collected in the cross-sectional survey. In our proposed approach, we adopt the current duration method to estimate the distribution of TBS using the available TSLS data, from which the frequency of sex at the population level can be derived. Furthermore, the observed TSLS data are subject to reporting issues because they can be reported in different units and may be rounded off. To apply the current duration approach and account for these data reporting issues, we develop a flexible Bayesian model, %We examine our proposed approach through simulation under various settings and validate our model. We apply the proposed approach to the TSLS data from Demographic and Health Surveys in African countries. 
and provide a detailed technical description of the proposed modeling approach.
\\Keywords: Bayesian inference; current duration approach; Demographic and Health Survey; family planning; reporting issues; sexual activity
\end{abstract}

\clearpage 
\section{Introduction}

Aggregate measures of family planning are used to monitor demand for and usage of contraceptive methods in populations globally, for example as part of the FP2030 initiative\footnote{See \url{https://fp2030.org/}}. Family planning measures for low- and middle-income countries are typically based on data collected through cross-sectional household surveys, such as the Demographic and Health Survey program \cite{DHS2018}. Recently proposed measures account for sexual activity through assessment of the distribution of time-between-sex (TBS) in the population of interest \cite{Leeetal2023}. In this paper, we introduce our statistical approach to estimate the distribution of TBS in order to comprehensively quantify women's sexual activity in the population.

There are some challenges in estimating the TBS distribution in our application of interest. First, the TBS data are not directly available for the populations of interest. Instead, time-since-last-sex (TSLS) data are collected from cross-sectional surveys, where participants are asked ``when was the last time you had sexual intercourse?'', which only provide limited information about women's sexual activity. The TSLS, or the \textit{current duration}, is the duration of time from the most recent sexual intercourse to the time of survey. In our proposed approach, we apply the current duration approach to make inferences about the unobserved TBS using the observed TSLS data \cite{Keidingetal2002}. 
In addition, TSLS data are subject to multiple reporting issues. In cross-sectional surveys, women may report their TSLS in different units such as days, weeks, months, and years. Also, respondents may prefer reporting TSLS on certain days, e.g., the multiples of 7 or 30. Hence, the data consist of a mix of exactly reported values and some coarse values, leading to ``heaped'' data. We apply a Bayesian modeling approach to address these reporting issues.
    
%TO DO: how our work fits into the literature (prob more than 1 para): what exists and what do we add? Brief introduction to the current duration approach, that we need to estimate the distribution of time-since-last-sex as a decreasing function. that we contribute to lit by new bayesian semi-parametric approach to do so, with set up motivated by and accounting for data reporting issues 

%This paper is organized as follows.... [finalize at end] %In section~\ref{sec-meth}, we propose a new nonparametric Bayesian approach to estimate TSLS distribution. We assess the performance of the proposed method through simulation studies in section~\ref{sec-sims}. We use DHS data collected from 4 African countries to estimate TSLS distribution for each country as case studies in section~\ref{sec-case}.

In this paper, we provide a detailed technical description of the modeling approach used to estimate the TBS densities from the TSLS data using a flexible Bayesian model. We introduce the notation and the model in Section 2, followed by a brief introduction of the relationship between the TBS and TSLS distributions based on the current duration approach in Section 3. In Section 4, we describe the proposed Bayesian estimation approach, by which the reporting issues in cross-sectional survey data are dealt with. We conclude by providing information on computation.

\section{Notation and summary of approach}
Let $X$ denote the discrete TBS in days for a woman randomly sampled from the population of interest. When the discrete time is $d$ days, i.e., $X = d$, we assume that the underlying (continuous) time between sex lies in the interval $(d, d+1]$. For example, $X=0$ refers to the time between sex up to 24 hours. Our goal is to estimate the distribution of TBS, $f_X(x) = \text{Prob}(X=x)$ with $X \sim  f_X(x)$, which implies the frequency of sex at the population level. 

In our application of interest, TBS data are not directly observed. Therefore, we utilize the TSLS data that are collected from women in the population. Assume we randomly sampled $n$ numbers of women from the population. For the $i$th woman, let $y_i$ denote the exact values of TSLS in days for $i=1, 2, \hdots, n$. We assume that $y_i \sim f_Y(y)$, where $f_Y(y) = \text{Prob}(Y=y)$ refers to the distribution of TSLS. The distribution of TBS, $f_X(x)$, can be estimated using the estimated TSLS distribution based on assumptions regarding the frequency of sexual intercourse. Details on the relationship between $f_X(x)$ and $f_Y(y)$ are given in Section~\ref{sec-cdapproach}. %In summary, we assume that sexual intercourse in the population does not have any systematic pattern, under which a relation between $f_X(x)$ and $f_Y(y)$ follows, constraining $f_Y(y)$ to be a decreasing distribution. 
Section~\ref{sec-splines} introduces the flexible Bayesian model used to estimate the TSLS distribution in this context.

The TSLS data may be reported in different units. For the $i$th woman, the observed data consist of $(z_i, u[i])$, where $z_i$ denotes the reported value of TSLS and $u[i]$ indicates the reporting unit with $u[i] = 1, 2, 3, 4$ referring to days, weeks, months, and years, respectively. 
%TO ADD: with sampling weight $w[i]$ and cluster indicator $c[i]$.    
We estimate the distribution of TSLS, $f_Y(y)$, using the observed data $\bm{y}=\{(z_i, u[i]), i=1,\ldots,n\}$, accounting for reporting errors, as described in Section~\ref{sec-like}.

% We introduce the proposed Bayesian semiparametric model for $f_Y(y)$ in Section~\ref{sec-splines}, and provide details on how we account for reporting issues in Section~\ref{sec-like}. 

\section{Relation between TBS and TSLS distributions }\label{sec-cdapproach}
%explain how we obtain the distribution of TBS from TSLS, cite relevant literature, summarize assumptions made. 
Assume women are sampled randomly from the population of interest, and for each woman, a gap time between two sexual intercourse events is sampled randomly from the gap times between the woman's sexual intercourse events. 
% Then, %a gap time of 10 days, for example, is 10 times more likely to be sampled than a shorter gap time of 1 day. In other words, 
% at the time of survey, when recording the TSLS, this reporting is more likely to capture gap times between sex that are longer than those with shorter duration. 
Also, assume that there is no systematic pattern in women's sexual intercourse events in the population. %Then, the probabilities of observing TSLS equal to $0, 1, \ldots, 10$ days at the time of survey are equally distributed for a gap time between sex that is equal to 10 days. %That is, conditional on $x=10$, $y$ is uniformly distributed at $0,\ldots,10$ days.
%We assume that women have sexual intercourse according to a homogeneous Poisson process with intensity $\lambda$. 
Under these assumptions, 
%The homogeneous Poisson process assumption introduces 
the following relation between $f_X(x)$ and $f_Y(y)$ follows \cite{Allison1985, Keidingetal2002}:
\begin{equation}\label{eq:1}
    f_Y(y) = \frac{S_X(y)}{E_X(X)},   
\end{equation}
where $S_X(x)=\text{Prob}(X\geq x)$ refers to the survival function associated with $X$ and $E_X(X)$ is the expected value of $X$. By Equation~(\ref{eq:1}), $f_Y(y)$ is constrained to be a decreasing distribution. Given $f_Y(y)$, the discrete distribution $f_X(x)$ can be derived as follows: 
\begin{equation*}\label{eq:2}
    f_X(x) = S_X(x)-S_X(x+1)=\frac{f_Y(x)}{f_Y(0)}-\frac{f_Y(x+1)}{f_Y(0)},   
\end{equation*}
where $f_Y(0)=1/E_X(X)$.

In applications, it is common to specify a fixed constant value $y_0<\tau$, where $\tau<\infty$ is the maximum observed $y$, to restrict the analysis to $y$ that satisfies $0\leq y<y_0$ excluding any duration that is longer than $y_0$. This ensures the quality of the data and/or refines the population of interest. For studying sexual activity, we excluded those who did not have sexual intercourse within the past 2 years. We restricted the support of $y$ to $0\leq y<y_0$ by placing an additional constraint $f_Y(y_0)=0$ or, equivalently $S_X(y_0)=0$, during the estimation of $f_Y(y)$ for $0\leq y <y_0$, which guarantees the derivation of an unrestricted TBS distribution, $f_X(x)$ over its support of $0\leq x <y_0$. Setting $S_X(y_0)=0$ is appropriate for our application where the empirical survival curve for TSLS is close to 0 at 2 years.

\section{Flexible Bayesian specification of TSLS distribution }\label{sec-splines}
%summarize relevant literature for Bayesian estimation of a decreasing distribution. point out that the non-parametric approach based on DP is not suitable here due to how data are reported. we propose a semi-parametric approach.   
We estimate the non-increasing discrete distribution $f_Y(y)$ using a flexible Bayesian model. For ease of notation, let $\phi_d = f_Y(d) = \text{Prob}(Y = d)$ denote the probability that the TSLS equals $d$ days. By definition, the $\phi_d$'s are constrained to be between 0 and 1 and to sum to 1: $\sum_{d=0}^{D} \phi_d = 1$ where $D = 729$ days (i.e., 2 years). To incorporate these constraints, we define $\phi_d$ on a transformed scale. 
Specifically, we define vector $\bm{\phi} = (\phi_0, \phi_1, \hdots, \phi_{D})$ as a transform of the vector $\bm{\gamma} = (\gamma_0, \gamma_1, \dots, \gamma_{D})$ as follows: 
\begin{align*}
    \bm{\phi} = \bm{\gamma}/\sum_d \gamma_d.
\end{align*}
With this transformation, the vector $\bm{\phi}$ is a well defined set of probabilities (i.e., all are between 0 and 1, and the vector sums to one), as long as $\bm{\gamma}\geq 0$.

% old softmax
% Specifically, we define vector $\bm{\phi} = (\phi_0, \phi_1, \hdots, \phi_{364})$ as the softmax transform of vector $\bm{\gamma} = (\gamma_0, \gamma_1, \dots, \gamma_{364})$ as follows: 
% \begin{align*}
%     \bm{\phi} = \mathrm{softmax}(\bm{\gamma}),
% \end{align*}
% where 
% \begin{align*}
%     \mathrm{softmax}(\bm{\gamma}) = \left( \frac{\exp(\gamma_0)}{\sum_i \exp(\gamma_i)}, \frac{\exp(\gamma_1)}{\sum_i \exp(\gamma_i)}, \dots, \frac{\exp(\gamma_{364})}{\sum_i \exp(\gamma_i)} \right).
% \end{align*}
% %The softmax function maps $\bm{\gamma}$ to a new vector with values in $(0, 1)$ whose entries sum to 1.
% With this transformation, the vector $\bm{\phi}$ is a well defined set of probabilities (all are between 0 and 1, and the vector sums to one), while $\bm{\gamma}$ is unconstrained. 

\begin{figure}[htbp]
    \includegraphics[width=\textwidth]{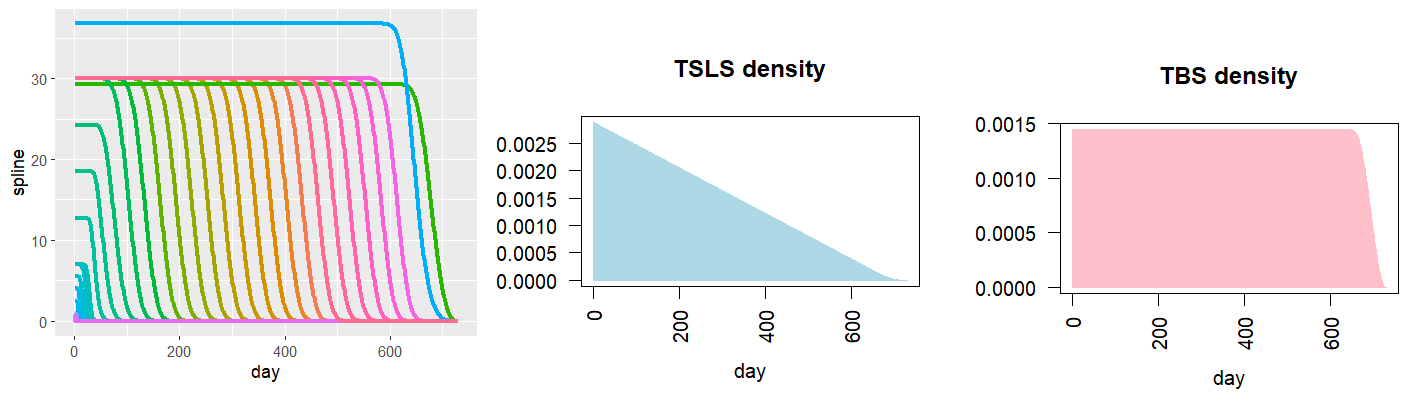}
  	\caption{\textbf{Illustration of spline functions and the resulting TSLS and TBS distributions when the spline coefficients are set to $\alpha_k = 1$ for all $k$.} (A) Integrated B-splines used for estimating the transformed TSLS probabilities. Each colored line represents one spline function $BI(\cdot)$, plotted against days. (B) Associated TSLS distribution when $\alpha_k=1$. (C) Associated TBS distribution.} 	\label{fig-splines}
\end{figure}

We model $\bm{\gamma}$ using integrated B-splines \cite{ramsay1988ispline}. The model specification for $\gamma_d$ is as follows:
\begin{align*}
    \gamma_d = \sum_{k=1}^K \alpha_k BI_k(d),
\end{align*}
where $BI_k(d)$ refers to the $k$th integrated B-spline evaluated at day $d$ and  $\bm{\alpha} = (\alpha_1, \alpha_2, \dots, \alpha_K)$ are the spline coefficients. The integrated B-splines used are shown in Figure~\ref{fig-splines} (A). We model the $\alpha_k$'s as follows:
\begin{align*}
    \alpha_k &= \exp \left( \sum_{j=k}^K \delta_j\right),\\
    \delta_j|\sigma &\sim N(0, \sigma^2), \\
    \sigma &\sim N^+(0,1).
\end{align*}
In this setup, the resulting ${\gamma}_d$'s are positive and decreasing with $d$. Figure~\ref{fig-splines} (B) and (C) illustrate the TSLS and TBS distributions, respectively, when $\delta_j=0$ for all $j$, $\alpha_k = 1$ for all $k$. 

%\paragraph{Likelihood function}
\subsection{Reporting assumptions}\label{sec-like}
We observe the reported TSLS value, $z_i$, along with the reporting unit $u[i]=1,2,3,4$ referring to days, weeks, months, or years, respectively. For data reported in weeks, we assume for the exact values of TSLS, $y_i$: 
$$y_i \in \{z_i\cdot 7, \hdots, (z_i+1)\cdot 7 - 1\}, \text{ for } u[i] = 2,$$
and for data reported in months, we assume:
$$y_i \in \{z_i\cdot 30+1, \hdots, (z_i+1)\cdot 30\}, \text{ for } u[i] = 3.$$ For data reported in years, where the reported values of TSLS is $z_i$ = 1 specifically, we assume:
$$y_i \in \{365 - 31, \hdots, 729\}, \text{ for } u[i] = 4.$$
For data reported in days, we assume that respondents may preferably report in multiples of $7$ or $30$ days as well as some may exactly report in days that are non-multiples of $7$ or $30$. This assumption is accounted for as follows: 
\begin{align*}
    y_i &= z_i \text{ for } u[i] = 1 \text{ and } z_i \notin \{7, 14, 21, 28, 30, 60, 90\},\\
    y_i &\in \{z_i-2, \hdots, z_i +2\} \text{ for } u[i] = 1 \text{ and } z_i \in \{7, 14, 21, 28, 30, 60, 90\}.
\end{align*}

Based on these reporting assumptions, the likelihood function is summarized as follows:
\begin{align*}
    z_i &\sim f_Z^{(day)}(\cdot) \text{ for } u[i] = 1 \text{ (reporting in days)},\\
    z_i &\sim f_Z^{(week)}(\cdot) \text{ for } u[i] = 2 \text{ (reporting in weeks)},\\
    z_i &\sim f_Z^{(month)}(\cdot) \text{ for } u[i] = 3 \text{ (reporting in months)},\\
      z_i &\sim f_Z^{(year)}(\cdot) \text{ for } u[i] = 4, z_i=1 \text{ (reporting in years)},
\end{align*}
where
\begin{align*}
      f_Z^{(day)}(z) &= \begin{cases} f_Y(z) & \text{ for } z \notin \{7, 14, 21, 28, 30, 60, 90\},\\
        \sum_{y = z-2}^{z+2}f_Y(y) & \text{ for } z \in \{7, 14, 21, 28, 30, 60, 90\},\end{cases}\\
      f_Z^{(week)}(z) &= \sum_{y = z\cdot 7 }^{(z+1)\cdot 7 - 1} f_Y(y),\\
    f_Z^{(month)}(z) &= \sum_{y = z\cdot 30 +1}^{(z+1)\cdot 30} f_Y(y),\\
      f_Z^{(year)}(1) &= \sum_{y = 365-31}^{729} f_Y(y).
\end{align*}
When visualizing the reported data in a histogram, we assign equal probability mass to the set of $y_i$ associated with $z_i$.

\subsection{Computation}
%Model Eq.(\ref{eq-pm}) defines set-up for the process model. 
We have developed an \texttt{R} package available on GitHub at https://github.com/AlkemaLab/safpet. For model fitting, an HMC algorithm is employed to sample from the posterior distribution of the model parameters that define the TSLS distribution with the use of \texttt{R} and \texttt{Stan} \cite{carpenter_stan_2017,R2021,Stan2018}. 
By using the Bayesian estimation model, we obtain a set of posterior samples of all model parameters. The estimated TSLS and TBS distributions are obtained from the posterior samples of the spline coefficients $\alpha_k^{(s)}$ for $s=1,\hdots, S$, where $s$ denotes the posterior sample index. Specifically, each combination of posterior samples $(\alpha_1^{(s)}, \hdots, \alpha_K^{(s)})$ provides one TSLS and TBS linked distribution. %To report the daily probabilities, we may use the point estimate the median value of that set at each day. 

The \texttt{R} and \texttt{Stan} code was used to produce the results for our recent application \cite{Leeetal2023}. In this application, we ran four parallel chains with a total of 2,000 iterations in each chain. The first 1,000 iterations in each chain from the warm-up phase are discarded so that the resulting chains contain 1,000 samples each. We used standard diagnostic checks to check convergence and sampling efficiency. These checks were based on trace plots, the improved Rhat diagnostic using rank-normalized draws (\cite{mcmc_diag_GR}, \cite{vehtari_rank-normalization_2020}), and various calculations of effective sample size (ESS), including the bulk ESS and the tail ESS - giving the minimum of the effective sample sizes of the 5\% and 95\% quantiles. 

% For the application, summary measures such as the average sexual exposure at a single day, which is the inverse of the mean TBS, were calculated using the set of TBS distributions: we calculated one summary per TBS distribution and report the median value of the resulting set of summary values as the point estimate. We also reported uncertainty in the form of 50\%, 80\%, and 95\% credible intervals, where the percentage refers to the probability of the true value being within the intervals. 

%------
\newpage
\bibliographystyle{unsrt}
\bibliography{SA_measure}
\end{document}